\documentclass[aps,prl,reprint,superscriptaddress,preprintnumbers,longbibliography]{revtex4-2}

\usepackage{amsmath,amssymb,bm,mathrsfs,bbold,dsfont}
\usepackage{xfrac}
\usepackage[dvipsnames]{xcolor}
\usepackage{physics}
\usepackage{slashed}
\usepackage{multirow}
\usepackage{booktabs}
\usepackage{array}

\usepackage{epsfig}
\usepackage{psfrag}
\usepackage{tikz}
\usepackage{float}
\usepackage[caption=false]{subfig}  % Prevents conflict with revtex4-1
\PassOptionsToPackage{caption=false}{subfig}

\usepackage[sort&compress]{natbib}
\usepackage[colorlinks=true,
    linkcolor=red,
    citecolor=blue,
    urlcolor=blue,
    filecolor=blue,
    anchorcolor=blue,
    menucolor=blue,
    linktocpage=true,
    pdfproducer=medialab,
    pdfa=true
]{hyperref}
\usepackage{cleveref}

\usepackage{setspace}
\usepackage{relsize}
\usepackage{enumerate}
\usepackage{comment}
\usepackage[normalem]{ulem}
\usepackage{srcltx}
\usepackage{xcolor}
\usepackage{diagbox}
\usepackage{appendix}

\newcommand{\dgap}{\Delta_{\rm gap}}      
\newcommand{\dmax}{\Delta_{\rm max}}

\newcommand{\calT}{\mathcal{T}}   
\newcommand{\Dg}{\Delta_{\rm gap}}

\newcommand{\emphb}[1]{\textbf{\emph{#1}}}

\newcommand{\Eq}[1]{Eq.~\eqref{eq:#1}}

\newcommand{\Fig}[1]{Fig.~\ref{fig:#1}}
\newcommand{\Figs}[2]{Figs.~\ref{fig:#1} and \ref{fig:#2}}
\newcommand{\Reff}[1]{Ref.~\cite{#1}}
\newcommand{\Reffs}[1]{Refs.~\cite{#1}}

\newcommand{\mysection}[1]{ \textbf{\textit{#1 ---}}}
\newcommand{\mysubsection}[1]{\textit{#1 ---}}

\begin{document}

\preprint{MIT-CTP/6144}

\newcommand{\ourtitle}{\boldmath Constraining the spectral gap of 2d CFTs just above $c=1$}

\title{\ourtitle}

\author{Nathan Benjamin}
\email{nathanbe@usc.edu}
\affiliation{Department of Physics and Astronomy, University of Southern California, Los Angeles, CA 90089, USA}

\author{Kevin Langhoff}
\email{langhoff@mit.edu}
\affiliation{Center for Theoretical Physics -- a Leinweber Institute, Massachusetts Institute of Technology, Cambridge, MA 02139, USA}

\author{Wei Li}
\email{weilibu17@gmail.com}
\affiliation{Department of Physics, Yale University, New Haven, CT 06511, USA}
\affiliation{Department of Physics, Boston University, Boston, MA 02215, USA}

\author{Jesse Thaler}
\email{jthaler@mit.edu}
\affiliation{Center for Theoretical Physics -- a Leinweber Institute, Massachusetts Institute of Technology, Cambridge, MA 02139, USA}
\affiliation{Institut des Hautes \'Etudes Scientifiques, 91440 Bures-sur-Yvette, France}
\affiliation{Institut de Physique Th\'eorique, CEA Paris-Saclay, 91191 Gif-sur-Yvette, France}
\affiliation{The NSF Institute for Artificial Intelligence and Fundamental Interactions}

%%%%%%%%%%%%%%%%%%%%%%%%%%%%%%%%%%%%%%%%%%%%%%%%%%%%
\begin{abstract}
We numerically study the space of compact unitary two-dimensional conformal field theories (2d CFTs) with central charge $c>1$ using the modular bootstrap.
We focus on bounding the spectral gap $\Delta_{\rm gap}$ of such theories, namely the smallest scaling dimension among the non-vacuum primary operators.
We use both a primal method that directly searches for consistent spectra and a dual method that yields rigorous exclusions.
A key feature of our analysis is requiring the degeneracies of primary operators to be positive integers.
At $c$ very near $1$, the integrality constraint greatly reduces the allowed spectral gap in both methods. For example, at $c-1 = 3\times 10^{-3}$ integrality strengthens the rigorous bound from $\Delta_{\rm gap} \lesssim 0.5$ to $\Delta_{\rm gap} < 0.041$.
Our results suggest that the spectral gap might vanish as $c$ approaches $1$ from above.
\end{abstract}
%%%%%%%%%%%%%%%%%%%%%%%%%%%%%%%%%%%%%%%%%%%%%%%%%%%%

\maketitle

%%%%%%%%%%%%%%%%%%%%%%%%%%%%%%%%%%%%%%%%%%%%%%%%%%%%
% Introduction
%%%%%%%%%%%%%%%%%%%%%%%%%%%%%%%%%%%%%%%%%%%%%%%%%%%%

\mysection{Introduction}
The space of two-dimensional conformal field theories (2d CFTs) is of great interest but remains poorly understood.
For central charge $c \leq 1$, compact unitary 2d CFTs have been fully classified \cite{Belavin:1984vu, Friedan:1983xq, Goddard:1986ee, Cappelli:1986hf, Kato:1987td, Ginsparg:1987eb, Dijkgraaf:1987vp, Eberhardt:2026qqm}.
For $c>1$, however, no such complete classification is known.
Even how close to $c=1$ a compact unitary CFT can be is an open question, with the smallest known central charge with $c>1$ coming from the $\mathbb{Z}_5$ parafermion CFT with $c = 8/7 \approx 1.14$~\cite{Fateev:1985mm}.
In this work, we use modular invariance to study theories with $c$ just above 1, with the goal of understanding what their properties might be or even whether they exist at all.

Modular invariance places strong constraints on the spectra of 2d CFTs \cite{Cardy:1986ie,Hellerman:2009bu}.
Conventional modular-bootstrap bounds typically assume a non-negative spectral density but ignore the requirement that the degeneracy of each primary operator must be an integer~\cite{Hellerman:2009bu,Friedan:2012jk,Keller:2012mr,Qualls:2013eha,Friedan:2013cba,Benjamin:2016fhe,Collier:2016cls,Bae:2017kcl,Dyer:2017rul,Afkhami-Jeddi:2019zci,Mukhametzhanov:2019pzy,Hartman:2019pcd,Benjamin:2019stq,Pal:2019zzr,Afkhami-Jeddi:2020hde,Collier:2021ngi,Benjamin:2022pnx,Benjamin:2025kvm,Rattazzi:2008pe,Poland:2018epd,Rychkov:2023wsd}.
Imposing integrality brings the bootstrap-derived spectra closer to that of an actual compact CFT.
A key question about the spectrum of 2d CFTs is the allowed spectral gap $\Delta_{\rm gap}$, namely the scaling dimension of the lowest non-vacuum primary.
As a preview, we find that integrality substantially constrains $\Delta_{\rm gap}$ as $c \to 1$ from above, with evidence that the spectral gap might almost completely close in this limit.

We investigate the impact of integrality on 2d CFTs using two complementary modular-bootstrap methods, building on \Reffs{Benjamin:2026lbj,Fitzpatrick:2023lvh}.
The \emph{primal} method searches directly for discrete spectra with integer degeneracies that approximately satisfy modular invariance.
The \emph{dual} method uses linear functionals to bound the number of primaries in a chosen range of dimensions; if those bounds leave no possible integer count, the assumed spectrum is excluded.
The terms primal and dual are used by analogy with duality in convex optimization.

These two methods are complementary.
The dual method gives rigorous exclusions, but it is computationally more expensive and cannot fully impose the discrete nature of primary operators. 
The primal method strictly enforces discreteness and integrality of primary operators, but interpreting the truncated solutions obtained from optimization requires care.
In particular, the primal method can neither rigorously exclude nor definitely prove the existence of a CFT.
Nevertheless, we find that the primal bounds are consistent with the dual ones once assumptions on the spectrum are appropriately relaxed.

\begin{figure*}[t]
    \centering
    \includegraphics[width=\linewidth]{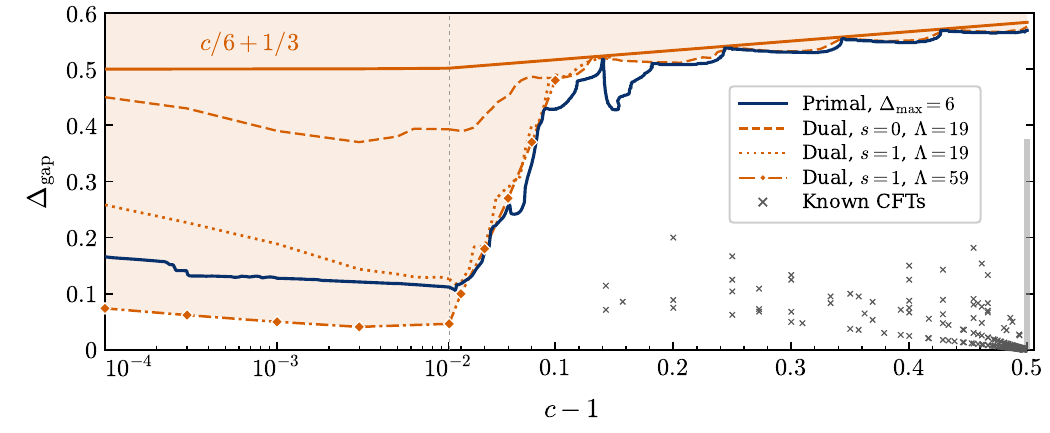}
    \caption{Constraints on the 
    spectrum of parity-symmetric 2d CFTs, showing the allowed spectral gap $\Delta_{\rm gap}$ as a function of the central charge $c$.
    Shaded orange regions are rigorously excluded.
    The solid orange line is the non-integral bound of $\Delta_{\rm gap} \leq c/6+1/3$~\cite{Collier:2016cls}.
    The dashed (dotted) orange line comes from the dual method with $\Lambda = 19$, obtained by requiring that the number of scalars (spin-$1$ parity pairs) with scaling dimension in the window $[\dgap,\dgap+w]$ ($[1,1+w]$) be an integer.
    The dot-dashed orange line is the same spin-$1$ analysis but with $\Lambda = 59$, computed at the ten $c$ values marked with diamonds.
    The blue line shows the approximate numerical boundary obtained from the primal method with $\dmax = 6$.
    Below $c \approx 1.02$, the $\dmax = 6$ primal boundary is weaker than the $\Lambda=59$ dual bound, owing to the fact that the optimal dual windows extend to $\Delta\approx 9$--$16$, beyond the $\dmax = 6$ primal truncation.
    Gray crosses mark known isolated CFTs while the gray bar at $c=3/2$ marks a known continuous family (see the Supplemental Material).
    }
    \label{fig:All_Bounds}
\end{figure*} 

Our main result is plotted in \Fig{All_Bounds}, and we summarize our main conclusions as follows:
\begin{enumerate}
    \item With the dual method, imposing integrality on the spin-$0$ sector yields only modest improvements on the rigorous non-integral bound of $\dgap \le c/6 + 1/3$ \cite{Collier:2016cls}.
    That said, we are able to set stronger bounds than the previous scalar analysis from \Reff{Fitzpatrick:2023lvh}.
    \item Our most surprising result in the dual method comes from imposing integrality in the spin-$1$ sector.
    This leads to qualitatively stronger bounds for $c \lesssim 1.1$.
    For $c=1.003$, at our level of derivative expansion, we lower the rigorous bound from $\dgap < 0.5005$ without integrality to $\dgap < 0.041$. 
    \item For $c \gtrsim 1.2$, dual and primal methods often develop kinks at similar values of $c$ with bounds on $\Delta_{\rm gap}$ close to the non-integral bounds.
    \item For $c \lesssim 1.2$, the primal bounds exhibit additional strong kinks, often (though not always) at the location of the dual bounds.
    We interpret this as evidence that there might be further gains to be made by pushing the dual method with integrality.
\end{enumerate}

\mysection{Modular Bootstrap Review}
We focus on compact, unitary, parity-symmetric 2d CFTs with $c>1$ (see e.g.~\cite{Yin:2017yyn} for a review).
The Hilbert space is organized according to the Virasoro algebra and its anti-holomorphic copy.
Primary states $\ket{\Delta,s}$ are labeled by their scaling dimension $\Delta$ and spin $s$, where unitarity requires $\Delta \geq |s|$.
We assume parity symmetry throughout, such that specifying the spin magnitude $s\geq0$ always stands in for identical but opposite $\pm s$-spin pairs.

\mysubsection{Primary Spectral Density} 
It is convenient to discuss assumptions on the non-vacuum primary spectrum in terms of its \emph{primary spectral density} $\rho_s(\Delta)$.
For a compact CFT, it takes a discrete form with positive integer degeneracies $d_{\Delta,s}$:
\begin{equation}
 \rho_{s}(\Delta)= \sum_{\{\Delta',s'\}} d_{\Delta',s'}\,
 \delta(\Delta-\Delta') \, \delta_{s,s'},
 \label{eq:integer_density_setup}
\end{equation}
which is the starting point for our primal method.

For the dual bootstrap, we consider weaker conditions on $\rho_s(\Delta)$.
The weakest we consider is just non-negativity:  $\rho_s(\Delta)\geq0$.
The workhorse condition for our dual analysis is \emph{window integrality}.
Specifically, define a window $W$ as:
\begin{equation}
W(a,b; s_0) \equiv \{(\Delta,s):\Delta \in [a,b],\,s = s_0\},
\end{equation}
with associated indicator function $\mathbb{1}_W(\Delta,s)$ that equals one if $(\Delta,s) \in W$ and zero otherwise.
The number of scalars or parity pairs ($s_0>0$) in $W$ is:
\begin{align} \label{eq:intdensity}
    N_W \equiv \sum_{s\ge 0} \int d\Delta \, \rho_s(\Delta)\,\mathbb{1}_W(\Delta,s) .
\end{align}
Window integrality is the requirement that $N_W$ is a non-negative integer (possibly zero).

\mysubsection{Partition Functions}
The torus partition function $Z$ collects information about the Hilbert space $\mathcal H$:
\begin{align}
 Z(\tau,\bar\tau)=\operatorname{Tr}_{\mathcal H}
 \left(q^{L_0-c/24} \,\bar q^{\bar L_0-c/24}\right),
 \label{eq:Ztrace}
\end{align}
with $q=e^{2\pi i\tau}$ and $\bar q=e^{-2\pi i\bar\tau}$.
We work on a Euclidean torus where $\tau = \tau_1 + i \, \tau_2$ and $\bar{\tau}$ is the complex conjugate of $\tau$.
The state $\ket{\Delta,s}$ is an eigenvector of $L_0$ and $\bar{L}_0$ with respective eigenvalues:
\begin{equation}
h = \frac{\Delta+s}{2}, \qquad \bar{h} = \frac{\Delta-s}{2}.
\end{equation}

The contribution of a primary and its descendants to $Z$ is determined by its Virasoro character.
At $c>1$, the holomorphic characters are:
\begin{equation}
\label{eq:characters}
\chi_0(\tau)=\frac{q^{-\frac{c-1}{24}}(1-q)}{\eta(\tau)}, \qquad \chi_{h>0}(\tau)=\frac{q^{h-\frac{c-1}{24}}}{\eta(\tau)},
\end{equation}
where $\eta(\tau)$ is the Dedekind eta function.
The factor $1-q$ in $\chi_0$ removes the vacuum null descendant $L_{-1}|0\rangle$.
A non-chiral primary contributes $\chi_h\bar\chi_{\bar h}$ to $Z$, with the exchanged product added for an opposite-spin pair.

It is convenient to define a reduced partition function~\cite{Friedan:2013cba}
that removes the common descendant factor:
\begin{equation}
 \widehat Z(\tau,\bar\tau)=\sqrt{\tau_2}\, |\eta(\tau)|^2 \,Z(\tau,\bar\tau).
 \label{eq:Zhat}
\end{equation}
We denote the contribution to $\widehat{Z}$ from a scalar primary ($s=0$), an opposite-spin pair ($s>0$), or the vacuum by:
\begin{align}
K_{\Delta,s}(\tau,\bar\tau)={}&(2-\delta_{s,0})\sqrt{\tau_2}\, 
 e^{2\pi\tau_2\left(\frac{c-1}{12}-\Delta\right)}\,\cos(2\pi s\tau_1),
 \label{eq:chi_Delta_s_Simple}\\
 K_{\rm vac}(\tau,\bar\tau)={}&\sqrt{\tau_2}\,e^{2\pi\tau_2\frac{c-1}{12}}|1-q|^2.
 \label{eq:chi_vac_Simple}
\end{align}
In terms of the spectral density, $\widehat Z$ takes the form:
\begin{equation}
 \widehat Z(\tau,\bar\tau)=K_{\rm vac}(\tau,\bar\tau)
 +\sum_{s\geq0}\int d\Delta\,\rho_s(\Delta)K_{\Delta,s}(\tau,\bar\tau).
 \label{eq:Z_density}
\end{equation}

\mysubsection{Modular Invariance}
A key feature of a CFT is that its partition function is unchanged by modular transformations of the torus defined by $\tau$.
Invariance under $T:\tau\mapsto\tau+1$ requires integer spins.
Invariance under $S:\tau\mapsto-1/\tau$ yields the 
\emph{modular bootstrap equation}:
\begin{equation}
\label{eq:MBE}
 Z(\tau,\bar\tau) = Z(-1/\tau,-1/\bar\tau).
\end{equation}
Since $\sqrt{\tau_2}|\eta(\tau)|^2$ is a modular invariant, $\widehat Z$ in \Eq{Zhat} also obeys \Eq{MBE}.
Letting $F_X(\tau,\bar\tau) = K_X(\tau,\bar\tau) - K_X(-1/\tau,-1/\bar\tau)$ for any subscript $X$, the modular bootstrap equation can be written as:
\begin{equation}
 F_{\rm vac}(\tau,\bar\tau)
 +\sum_{s\geq0}\int d\Delta\,\rho_s(\Delta)F_{\Delta,s}(\tau,\bar\tau)=0.
 \label{eq:MBE_density}
\end{equation}

\mysection{Primal Method}
To frame the modular bootstrap equation as an optimization problem, we must specify a finite search space, a loss function to minimize, and a procedure for finding and assessing minima.
Truncating the spectrum and solving the crossing equations approximately has precedent in the bootstrap~\cite{Gliozzi:2013ysa,Gliozzi:2016cmg,Esterlis:2016psv,Niarchos:2023lot}, although those methods impose exact crossing on a finite set of operators rather than scoring an approximate residual~\cite{Benjamin:2026lbj}.

\mysubsection{Trial Spectrum}
At each retained spin, we introduce $n_s$ primary states with free scaling dimensions and unit degeneracy.
To impose a spectral gap, we fix the lightest scalar operator at $\dgap$.
Since we only consider $\dgap<1$, integer spin and unitarity ensure that the lightest primary is a scalar.
The truncation is set by a cutoff $\dmax$: a candidate spectrum keeps only primaries with $\Delta\leq\dmax$, though operators with $\Delta>\dmax$ are retained during optimization.

\mysubsection{Truncation Uncertainty}
After truncating at $\dmax$, even a true CFT will fail to exactly satisfy modular invariance.
To see the size of the expected mismatch, split the reduced partition function into retained and omitted contributions, $ \widehat Z=\widehat Z_{\leq}+\widehat Z_{>}$, where contribution from the vacuum and primaries with scaling dimension up to $\dmax$ are contained in $\widehat Z_{\leq}$.
The residual violation of modular invariance is set by the missing contribution:
\begin{align}
 R(\tau)&\equiv\widehat Z_{\leq}(\tau)-\widehat Z_{\leq}(-1/\tau) \nonumber\\&=-\bigl[\widehat Z_{>}(\tau)-\widehat Z_{>}(-1/\tau)\bigr].
 \label{eq:truncation_balance}
\end{align}
An absolute residual alone cannot tell us whether a candidate spectrum is comparably accurate to what one would expect from a true CFT unless the contributions from truncated primaries can be estimated.

Following \Reff{Benjamin:2026lbj}, we estimate the truncation uncertainty using the free boson on a circle of radius $\sqrt2$, deformed from $c=1$ to the central charge being tested.
We combine contributions of the truncated part of the spectrum for this reference CFT at $\tau$ and $-1/\tau$ in quadrature to obtain a characteristic truncation uncertainty:
\begin{equation}
 \sigma_{\rm trunc}(\tau)^2 \equiv
\widehat Z^{\rm ref}_{>}(i|\tau|)^2+\widehat Z^{\rm ref}_{>}(i/|\tau|)^2.
 \label{eq:reference_sigma}
\end{equation}
Evaluating at $i|\tau|$ rather than $\tau$ prevents oscillations from making $\sigma_{\rm trunc}$ accidentally small~\cite{Benjamin:2026lbj}.

\mysubsection{Loss Function}
We minimize a loss function written as the mean squared residual normalized by the truncation uncertainty:
\begin{align}
 \mathcal L=\frac{1}{|\calT|}\sum_{\tau\in\calT}
 \left(\frac{R_{\rm train}(\tau)}{\sigma_{\rm trunc}(\tau)}\right)^2,
 \label{eq:primal_loss}
\end{align}
where $\calT$ is a sample of modular parameters and $R_{\rm train}(\tau)=F_{\rm vac}(\tau)+\sum_{\Delta,s}d_{\Delta,s}F_{\Delta,s}(\tau)$, with the sum over the full trial spectrum, including primaries above $\dmax$ (Supplemental Material, \Eq{spectrum_param}).
In practice, $\sigma_{\rm trunc}$ is combined in quadrature with a floating-point rounding scale to regulate the minimization; at $\dmax=6$, this never decides whether a spectrum is admitted.

A squared normalized residual of order one means the residual is comparable to the estimated truncation uncertainty; this is as small as one could hope for, even for a true CFT.
A much larger loss means a candidate does not reproduce modular invariance to the estimated accuracy.

If the residual errors obtained from truncating a true CFT at different $\tau$ were purely statistical with uncertainty $\sigma(\tau)$, the distribution of $(R(\tau)/\sigma(\tau))^2$ over many $\tau$ would follow a $\chi^2$ distribution for a single degree of freedom. 
Truncation is not a statistical process, but truncated spectra of known CFTs do follow this distribution (\Fig{app:residuals_histogram}).
The loss can therefore be read quasi-statistically: $\mathcal{L}>25$ is very roughly a $5\sigma$ exclusion.
To obtain a more robust boundary, we use a separate evaluation loss involving only primaries with $\Delta < \dmax$  (\Eq{app:eval_loss}), though it makes no visible impact on the results.

\mysubsection{Minimization}
The characters depend exponentially on dimension, so the Jacobian of normalized residuals is badly conditioned.
We therefore minimize with singular value descent (SVEN)~\cite{Bright-Thonney:2026tlr}, a modified Gauss-Newton method that inverts the Jacobian only on the leading singular directions.  

\mysubsection{Primal Results}
The blue line in \Fig{All_Bounds} shows the $\mathcal L_{\rm eval}=25$ primal boundary for $\dmax = 6$; other $\dmax$ are shown in \Figs{app:landscapes_dmax}{app:dmax_frontier}.
The boundary is insensitive to the threshold, since the loss jumps by orders of magnitude across it (\Fig{Evaluation Loss}).
Spectra above this boundary are tentatively excluded, in the sense that no solution was found with small enough loss.
To obtain a rigorous bound, though, we must turn to the dual bootstrap.

\mysection{Dual Method}
Using just non-negativity of the spectral function, we can set an upper bound on $\dgap$ using the dual method.
At a fixed $c$ and proposed gap $\dgap$, we define a linear functional $\alpha$ mapping $F_{\rm vac}$ and $F_{\Delta,s}$ to real numbers.
Suppose it satisfies:
\begin{equation}
 \alpha[F_{\rm vac}]>0, \qquad
 \alpha[F_{\Delta,s}]\geq0,~\forall \Delta \geq \dgap.
 \label{eq:positive_functional}
\end{equation}
Applying $\alpha$ to \Eq{MBE_density} implies the sum can not vanish. 
The numerical bound from non-negativity was found to extrapolate to $\dgap\leq c/6+1/3$ for $1\leq c\leq4$~\cite{Collier:2016cls}, improving earlier functional bounds~\cite{Hellerman:2009bu,Friedan:2013cba}.
The above argument only assumes that spectral density is non-negative and ignores the discrete spectrum of primary operators.

\mysubsection{Window Integrality}
The method of \Reff{Fitzpatrick:2023lvh} adds information about integrality by bounding $N_W$ from \Eq{intdensity}.
A proposed spectrum is excluded when the resulting range of $N_W$ contains no integer.

Given a window $W$, suppose we have two linear functionals $(\alpha_-,\, \alpha_+)$ that satisfy:
\begin{align}
 \alpha_-[F_{\Delta,s}]\leq\mathbb{1}_W(\Delta,s)
 \leq\alpha_+[F_{\Delta,s}],~\forall\Delta \geq \dgap.
 \label{eq:count_functionals}
\end{align}
This implies that $N_W$ is bounded as:
\begin{align}
 \sum_{s\ge 0}\int d\Delta\,\rho_s\,\alpha_-[F_{\Delta,s}]
 \leq N_W\leq \sum_{s\ge 0}\int d\Delta\,\rho_s\,\alpha_+[F_{\Delta,s}].
\end{align}
Modular invariance replaces each integral by minus the functional acting on the $F_{\rm vac}$, such that:
\begin{align}
 N_W^-\leq N_W\leq N_W^+,
 \qquad N_W^\pm=-\alpha_\pm[F_{\rm vac}].
 \label{eq:count_bounds}
\end{align}
At fixed $c$, $\dgap$, and $W$, we can separately vary the coefficients of $\alpha_\pm$ to minimize (maximize) $-\alpha_\pm[F_{\rm vac}]$, subject to the inequalities in \Eq{count_functionals}. 
For a compact CFT, the resulting interval must admit an integer count, such that the proposed gap is excluded if:
\begin{align}
     [N_W^-,N_W^+]\cap\mathbb Z_{\geq0}=\varnothing.
 \label{eq:integer_exclusion}
\end{align}

Importantly, bounds from window integrality depend on the specific window chosen.
We consider windows for $s=0$ primaries with $\Delta \in [\dgap,\dgap+w]$ and separately for $s=1$ primary pairs with $\Delta \in [1,1+w]$.
Windows for $s\geq 2$ are almost always less constraining than $s=1$.
Considering multiple windows over multiple spins simultaneously may lead to stronger bounds, but so far, we find that they add little constraining power.
At each proposed $\dgap$, we first optimize the two functionals for a fixed width $w$ and test the resulting count interval for an integer.
We then vary $w$ and repeat until we optimize bounds on $\dgap$.
At $c$ very close to $1$, we find the strongest bounds for $s=1$; we give a heuristic explanation for this in the Supplemental Material.

\mysubsection{Optimizing Functionals}
We optimize over functionals spanned by~\cite{Collier:2016cls}:
\begin{align}
 \alpha=\sum_{\substack{ k+\bar k\ {\rm odd}\\ k+\bar k\leq\Lambda}}\lambda_{k\bar k}\,
 \partial_\tau^{k}\partial_{\bar\tau}^{\bar k}\Big|_{\tau=i},
 \label{eq:functional_basis}
\end{align}
where $\Lambda$ is the cutoff and only $k+\bar k\ {\rm odd}$ survives on functions that are odd under $S$.
Crucially, the cutoff $\Lambda$ \emph{does not} change the rigor of any bounds obtained through the dual method, only their strength.
Acting on $F_{\Delta,s}$, every basis element returns an exponential times a polynomial:
\begin{align}
 \alpha[F_{\Delta,s}]=e^{-2\pi\left(\Delta-\frac{c-1}{12}\right)}p_s(\Delta),
 \label{eq:exp_poly}
\end{align}
with $p_s$ linear in the $\lambda_{k\bar k}$ and of degree $\Lambda$.

Outside a window $W$ at spin $s_0$, \Eq{count_functionals} requires that $\pm p_s\geq0$ for $\alpha_\pm$ on a half line $\Delta\geq\Delta_s^0$, with $\Delta_s^0=\max(\dgap,s)$ for $s\neq s_0$ and $\Delta_s^0$ the upper edge of $W$ for $s=s_0$.
This is imposed exactly by the sum-of-squares form~\cite{Rattazzi:2008pe,Poland:2018epd}:
\begin{align}
\pm p_s(\Delta)=\bm v^{T}Q\,\bm v+(\Delta-\Delta_s^0)\,\bm u^{T}H\,\bm u,\quad Q,H\succeq0,
 \label{eq:sos}
\end{align}
with $\bm v,\bm u$ vectors of monomials in $\Delta$.
In this way, $\alpha_\pm$ becomes a semidefinite program, which we optimized using SDPB~\cite{Simmons-Duffin:2015qma,Landry:2019qug} under the constraint $\pm \alpha_\pm[F_{\Delta,s_0}]\geq\pm 1$ imposed inside $W$.

\begin{figure}[t]
    \centering
    \includegraphics[width=\linewidth]{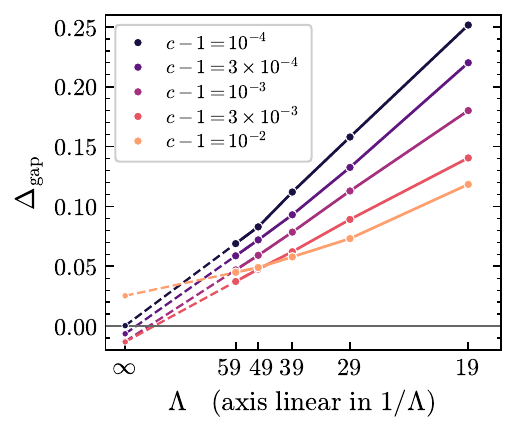}
    \caption{Window-optimized gap upper bound $\Delta_{\rm gap}$ versus derivative order $\Lambda$ (axis linear in $1/\Lambda$) for five values of $c-1$.
    Dashed lines pass through the $\Lambda=49$ and $59$ points and are continued to $\Lambda\to\infty$.
    The lines only guide the eye; three of the five cross zero before $\Lambda\to\infty$ and a fourth ends just above it, so the sequences must flatten out as no bound can fall below $(c-1)/12$.}
    \label{fig:lambda_extrapolation}
\end{figure}

\mysubsection{Cutoff Dependence}
Since functionals of order $\Lambda$ are contained in those of order $\Lambda'>\Lambda$, raising $\Lambda$ only strengthens a bound for fixed window:
\begin{align}
 N_W^-(\Lambda)\leq N_W^-(\Lambda')\leq N_W\leq N_W^+(\Lambda')\leq N_W^+(\Lambda).
 \label{eq:nesting}
\end{align}
To obtain the strongest bound on the spectral gap, we optimize over the window width $w$:
\begin{equation}
\label{eq:corner}
\Delta_*(\Lambda)\equiv\min_{w}\,\dgap^{\rm excl}(\Lambda,w),
\end{equation}
which is also non-increasing in $\Lambda$.
We find that the optimal width $w_*$ grows in proportion to $\Lambda$ for $c-1\leq3\times10^{-3}$, while the excluded gap tracks the top of the window, so the bound falls roughly as $1/\Lambda$ for as long as the window keeps opening.
At $c-1=10^{-2}$ the width already grows more slowly than $\Lambda$ and the descent flattens, as shown in \Fig{lambda_extrapolation}.
At smaller $c$, we do not yet see any evidence that $\Delta_*(\infty)$ will saturate, though there is an irreducible floor, as we now explain.

\mysubsection{Bound on the Bound}
The $c=1$ free boson's partition function (at any radius) is modular invariant, with a positive, integral spectrum.
This partition function can be reinterpreted as that of a $c>1$ CFT with no vacuum state~\cite{Friedan:2012jk,Friedan:2013cba,Collier:2016cls,Benjamin:2019stq}, such that the $c = 1$ vacuum becomes a scalar operator of the $c > 1$ theory with dimension $\frac{c-1}{12}$ (and all vertex operators are interpreted as having dimensions above $\frac{c-1}{12}$).
Even though this partition function, when interpreted as a $c>1$ theory, lacks a vacuum, it is still a valid modular bootstrap solution, and therefore must vanish when any functional acts on its states.
Therefore, $\frac{c-1}{12}$ sets the strongest obtainable bound on $\dgap$ from our dual method.

%(a bound on $\dgap$ below $\frac{c-1}{12}$ would be sign-definite for all states in the $c=1$ free boson example).

\mysubsection{Dual Results}
The rigorous dual bounds on $\dgap$ are collected in \Fig{All_Bounds}.
Away from $c=1$, the integrality constraints improve only modestly from $c/6+1/3$; for $c\gtrsim1.2$, the spin-0 dual curves run within a few hundredths of it, in qualitative agreement with the primal boundary.

The picture changes dramatically as $c$ approaches $1$.
For $c\lesssim 1.1$, the dual bounds on $\dgap$ are now set by the spin-1 windows and depend strongly on the cutoff $\Lambda$.
For $c\gtrsim 1.01$, the dual bounds seemed to have converged.
At smaller $c$, the true bound may be much lower than the $\Lambda=59$ result shown here, extrapolating from \Fig{lambda_extrapolation}.
Interestingly, the dual bounds for $c\lesssim 1.02$ dip below the primal boundary, since the optimal windows for $\Lambda=59$ extend beyond the $\dmax = 6$ truncation.

\mysection{Conclusions}
In this paper, we showed that the requirement of integrality greatly constrains the space of possible 2d CFTs near $c=1$.
Imposing window integrality through the dual method, we obtain rigorous bounds on the spectral gap $\dgap$, finding that spin-1 windows dramatically improve upon the non-integral bound when $c\lesssim 1.01$.
The primal method offers an independent construction that more accurately implements the assumption of a discrete spectrum.
Wherever the dual cutoff $\Lambda$ and primal truncation scale $\dmax$ are consistent, the two methods yield comparable results, sometimes with a near vanishing dual/primal gap.

The most intriguing features appear when the primal method exhibits kinks not (yet) seen in the dual approach.
This includes a tantalizing spike in the primal boundary just above $c = 1.14$.
As shown in the Supplemental Material, these kinks are strongly dependent on $\dmax$, suggesting that additional dual constraints might be obtainable by going beyond window integrality.

Our work here has assumed parity, as well as a lack of any additional higher-spin conserved currents.
We do not expect the results to qualitatively change if either of these assumptions is relaxed, but it may be interesting to repeat the analysis done here without them.
It would also be interesting to extend our analysis strategy to the conformal bootstrap of CFT correlators.

It is still an open question whether any compact unitary 2d CFT exists at $c$ slightly above $1$.
Our analysis greatly constrains the space of such theories and perhaps hints at theories to be discovered at its boundaries.

\mysubsection{Note Added}
While this paper was being completed, \Reffs{Poland:2026zyr,nakayama2026crossingsymmetryforbidsjust,kousvos2026virasorobootstrapcsim1desert} appeared, finding related bounds near $c=1$ using the conformal bootstrap.

\mysubsection{Acknowledgments}
We are extremely grateful to A.~L.~Fitzpatrick for collaboration at an early stage of this project.
We thank
S.~Collier,
L.~Eberhardt,
R.~Erramilli,
M.~Ferragatta,
H.~T.~Lam,
J.~Parra-Martinez,
E.~Perlmutter,
L.~Rastelli,
B.~van~Rees,
S.~Rychkov,
and
H.~Saleur for valuable discussions.
We thank S.~Collier, L.~Eberhardt, and N.~Levine for sharing unpublished work on similar results using integrality in the modular bootstrap.

NB is supported by the U.S.\ Department of Energy (DOE) Office of High Energy Physics (OHEP) under grant Contract Number DE-SC0011687.
KL and JT are supported by the DOE OHEP under grant Contract Number DE-SC0012567 and by Simons Foundation Investigator Awards 929241 (Thaler) and 929255 (Slatyer).
JT is additionally supported by the U.S.\ National Science Foundation (NSF) under Cooperative Agreement PHY-2019786 (The NSF AI Institute
for Artificial Intelligence and Fundamental Interactions, http://iaifi.org/) and by the MIT Generative AI Impact Consortium (MGAIC).
JT also thanks the Institut des Hautes \'Etudes Scientifiques (IHES) and the Institut de Physique Th\'eorique (IPhT) for providing an inspiring sabbatical environment to initiate this research.
WL is supported by the DOE Early Career Award DE-SC0025581, DOE OHEP under Award Number DE-SC0015845, the Sloan Foundation, and the Simons Collaboration on Confinement and QCD Strings.

\mysubsection{Code Availability}
The code and the data behind the figures are available at \url{https://github.com/Kevin-Langhoff/cft-gap}.

\mysubsection{Use of AI}
The authors used Claude Opus 5.5 to assist with coding for the numerics, and both Claude Opus 5.5 and ChatGPT-6 Astra to assist in cataloging known CFTs with $1 < c < \frac32$ in the Supplemental Material.

\bibliography{bibliography}

%%%%%%%%%%%%%%%%%%%%%%%%%%
% Supplemental Material %
%%%%%%%%%%%%%%%%%%%%%%%%%%

\onecolumngrid

\newpage

\begin{center}
   \textbf{\large SUPPLEMENTAL MATERIAL \\[.2cm] ``\ourtitle”}\\[.2cm]
  \vspace{0.05in}
  {Nathan Benjamin, Kevin Langhoff, Wei Li, and Jesse Thaler}
\end{center}

\setcounter{equation}{0}
\setcounter{figure}{0}
\setcounter{table}{0}
\setcounter{section}{0}
\setcounter{page}{1}
\makeatletter

\renewcommand{\theequation}{S\arabic{equation}}
\renewcommand{\thefigure}{S\arabic{figure}}
\renewcommand{\thetable}{S\arabic{table}}
\renewcommand{\thesection}{S\Roman{section}}
\setcounter{equation}{0}
\renewcommand{\theHfigure}{S\arabic{figure}}
\renewcommand{\theHequation}{S\arabic{equation}}
\renewcommand{\theHtable}{S\arabic{table}}
\renewcommand{\theHsection}{S\Roman{section}} 

%%%%%%%%%%%%%%%%%%%%%%%%%%

\section{Primal Method Review: Searching for Approximate 2D CFT Spectra}

In this section, we give additional information about the construction of the primal method where we search over the space of truncated discrete 2d CFT spectra with integer degeneracies for approximate solutions to the modular bootstrap equation.
We describe (a) how the loss function can be interpreted and how this aids in understanding the relative quality of a solution obtained via optimization, (b) details on how the optimization is actually performed, and (c) how the landscape of CFTs over central charge $c$ and spectral gap $\dgap$ is explored.

\emphb{Trial Spectrum.} Scaling dimensions of primaries other than the scalar at $\dgap$ are parametrized by $x\in \mathbb{R}$: 
\begin{equation} \label{eq:spectrum_param}  \Delta=\Delta_s^-+  \frac{\Delta_s^+-\Delta_s^-}{1+e^{-x}}, \qquad  \Delta_s^-=\max(\dgap,s),\qquad  \Delta_s^+=2\dmax-\Delta_s^-. 
\end{equation}
The upper limit extends beyond $\dmax$. During optimization, all trial operators contribute to the loss, including those above $\dmax$.
We truncate only when evaluating the final candidate spectrum.
An operator can therefore move through the eventual cutoff while continuing to receive a gradient.
This allows the number of retained operators to change without changing the number of optimization variables. 
Empirically, increasing $n_s$ beyond some value no longer changes the number of operators kept below $\dmax$; excess primaries float above the cutoff and are truncated.

\emphb{The Loss Function: Calibrating with Known Spectra.}
To test the loss function normalization independently of optimization, we consider the known primary spectrum of the $\mathcal N=1$ minimal model with $m=5$, truncate it, and evaluate its residual in the modular bootstrap equation on the fixed sample of $952$ modular parameters used for all reported losses, following~\Reff{Benjamin:2026lbj}.

\begin{figure}[b]
    \centering
    \includegraphics[width=1\linewidth]{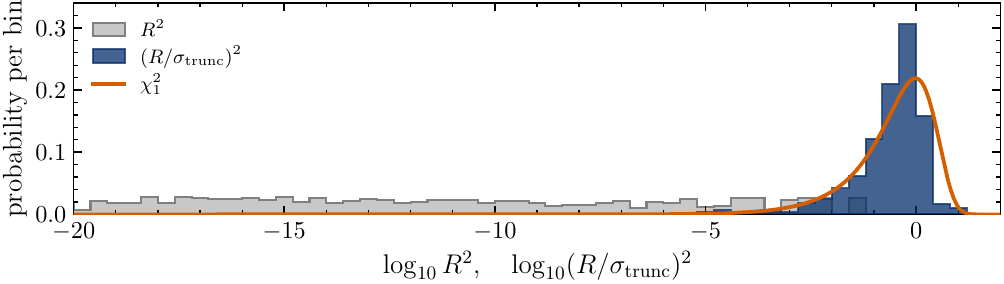}
    \caption{Distribution of squared residuals for a known CFT ($\mathcal{N}=1$ Minimal Model with $m = 5$) truncated at $\dmax=4$. The squared residual $R^2$ (gray) is spread over twenty decades, while the normalized square residual $(R/\sigma_{\rm trunc})^2$ (blue) concentrates near unity and approximately follows a $\chi^2$ distribution with one degree of freedom (orange). This is the quasi-statistical behavior that motivates the loss function of \Eq{primal_loss}.}
\label{fig:app:residuals_histogram}
\end{figure}

Figure~\ref{fig:app:residuals_histogram} shows the two motivations for the loss function of \Eq{primal_loss}.
For a known spectrum, the bare residual $R^2$ spans twenty orders of magnitude across the sample, while $(R/\sigma_{\rm trunc})^2$ concentrates near unity with a tail resembling a $\chi^2_1$ distribution.
The normalization turns the residual into a quantity whose natural size is one, which is what makes a loss of order one the floor for a truncated CFT and $\mathcal L=25$ a meaningful margin above it; read as a $\chi^2_1$ variable, $\mathcal L>25$ would very roughly be a $5\sigma$ exclusion.
The truncation error is not purely statistical across $\tau$, so the $\chi^2_1$ curve describes the distribution but does not strictly turn the threshold into a confidence level.

\begin{figure}[t]
    \centering
    \includegraphics[width=\linewidth]{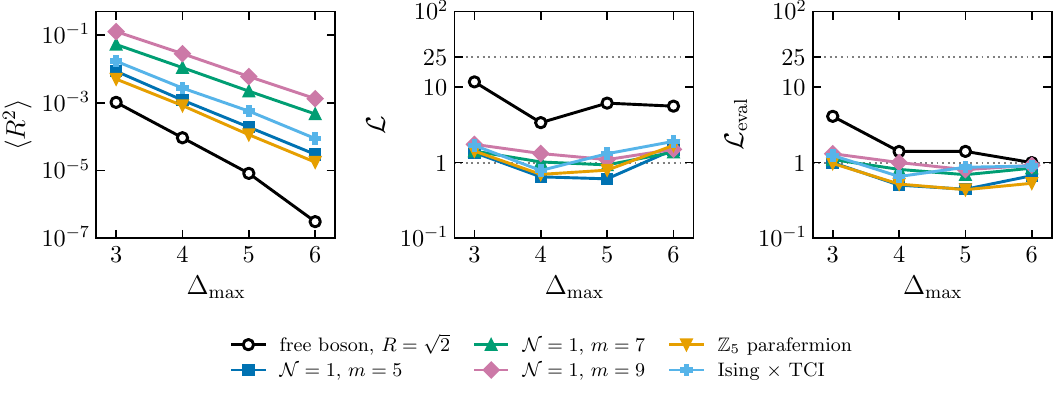}
    \caption{Mean squared residual (left), loss $\mathcal L$ (middle), and evaluation loss $\mathcal L_{\rm eval}$ (right) of known CFT spectra truncated at $\dmax=3$, $4$, $5$, and $6$, without any optimization.
    The bare residual falls exponentially with $\dmax$.
    Normalizing by $\sigma_{\rm trunc}$ removes this dependence, but $\mathcal L$ can still be dominated by a few modular parameters where $\sigma_{\rm trunc}$ is smallest, as it is for the free boson.
    By contrast, $\mathcal L_{\rm eval}$ is insensitive to these outliers and stays of order one for every theory.
    Dotted lines mark $1$ and the admission threshold $25$.
    The free boson is the reference spectrum that defines $\sigma_{\rm trunc}$.}
    \label{fig:app:cutoff_triple}
\end{figure}

Figure~\ref{fig:app:cutoff_triple} repeats the calibration for six known spectra at $\dmax=3$, $4$, $5$, and $6$.
Writing $r(\tau) = R(\tau)/\sigma_{\rm trunc}(\tau)$, we define the evaluation loss as:
\begin{equation}
\label{eq:app:eval_loss}
 \mathcal L_{\rm eval}=\exp\left[\frac{1}{|\calT|} \sum_{\tau} \log\left(1+r(\tau)^2\right)\right]-1 \leq \frac{1}{|\calT|} \sum_{\tau} r(\tau)^2= \mathcal L.
\end{equation}
This definition reduces the weight of a few large residuals.
The middle panel shows why this matters.
The free boson, which defines $\sigma_{\rm trunc}$, has the smallest bare residual of the six but the largest $\mathcal L$.
The loss $\mathcal L$ is dominated by points with the smallest $\tau_2$, where a spectrum as sparse as the free boson's is not capable of supplying a large enough cancellation in the modular bootstrap equation with a truncated spectrum.
This larger value of the loss is not necessarily a problem with the spectrum, but how certain values of $\tau$ can dominate the loss function purely due to finite truncation effects.
Suppressing such terms using $\mathcal{L}_{\rm eval}$ brings the free boson's loss to a more comparable value with the other known CFTs. 

\begin{figure}[b]
    \centering
    \includegraphics[width=\linewidth]{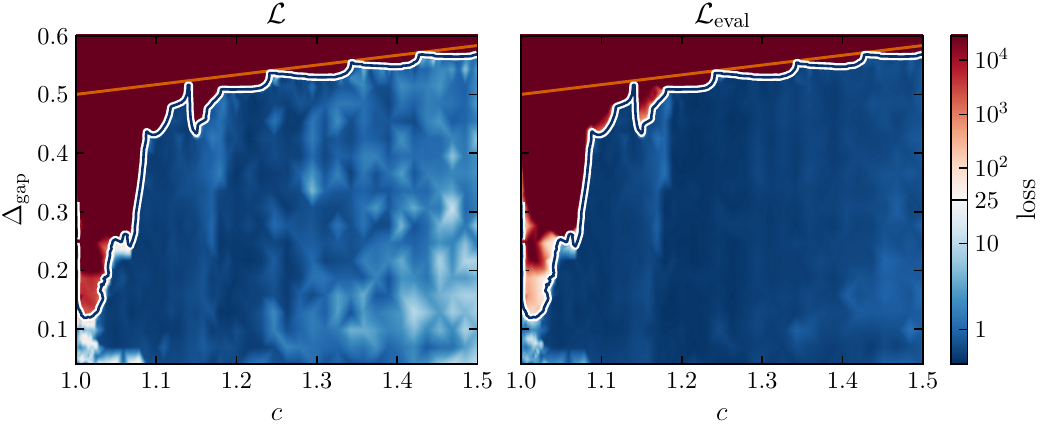}
    \caption{Loss $\mathcal L$ (left) and evaluation loss $\mathcal L_{\rm eval}$ (right) of the spectra found by the search at $\dmax=6$, as functions of $c$ and $\dgap$, on a common color scale with white at $25$. The contour $\mathcal L_{\rm eval}=25$ is drawn on both panels as a common reference; both losses rise by orders of magnitude across it. The orange line is $c/6+1/3$. The evaluation loss has the small advantage of removing some apparent noise at larger values of $c$. }
    \label{fig:Evaluation Loss}
\end{figure}

This choice of evaluation loss hardly affects the final boundary, as \Fig{Evaluation Loss} shows for the spectra found by the $\dmax=6$ search over the $(c,\dgap)$ plane.
Using $\mathcal{L}$, there are patches of large values inside the admitted region at larger $c$ where a few modular parameters dominate.
The evaluation loss $\mathcal L_{\rm eval}$ removes these outliers and leaves a uniform admitted region.
Both losses rise by several orders of magnitude across the contour $\mathcal L_{\rm eval}=25$, which is what makes the boundary insensitive to the chosen threshold.

\begin{figure}[t]
    \centering
    \includegraphics[width=\linewidth]{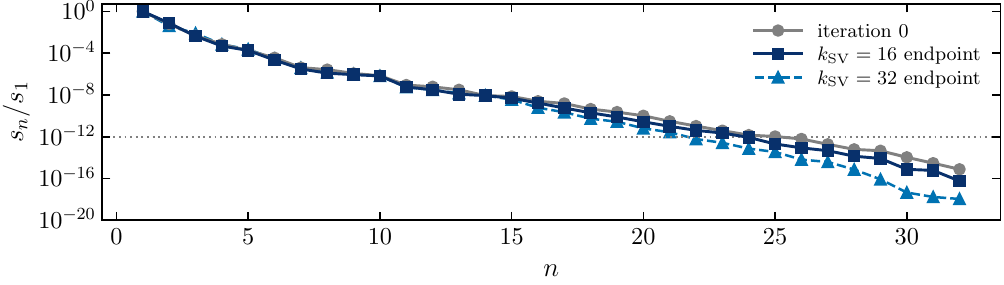}
    \caption{Singular values $s_n$ of the residual Jacobian, normalized to the largest, at the start of the optimization and at the endpoints of the $k_{\rm SV}=16$ and $32$ runs of \Fig{app:optimizer}. The spectrum falls exponentially with no plateau, so only the leading directions carry information about the residual, and directions below the solver cutoff $10^{-12}$ (horizontal dotted line) are discarded whatever $k_{\rm SV}$ is chosen.}
    \label{fig:app:singular_values}
\end{figure}

\begin{figure}[b]
    \centering
    \includegraphics[width=\linewidth]{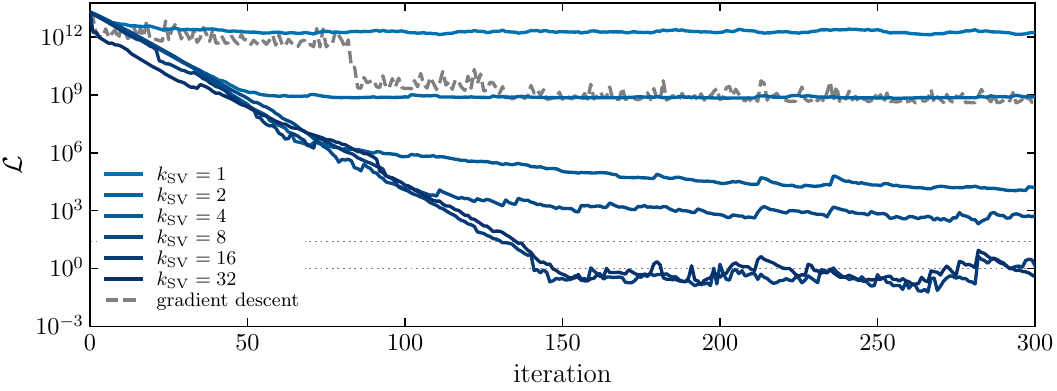}
    \caption{Loss $\mathcal L$ against iteration for gradient descent and for SVEN with $k_{\rm SV}$ retained singular directions, starting from the same initial spectrum at $\dmax=4$ and $(c,\dgap)=(81/70,3/35)$. Gradient descent and small $k_{\rm SV}$ stall many orders of magnitude above the admission threshold; from $k_{\rm SV}=16$ the loss reaches order one within $150$ iterations. Dotted lines mark $\mathcal L=1$ and $25$.}
    \label{fig:app:optimizer}
\end{figure}

\emphb{Minimization.} With the normalized residuals $r(\tau)$ and the search parameters $\theta_i$, a Gauss-Newton step solves for the Jacobian $J_{\tau i}=\partial r(\tau)/\partial\theta_i$.
Because every character depends exponentially on its dimension, the singular values of $J$ fall over ten decades across the parameters, as shown in \Fig{app:singular_values}, so a full inverse amplifies the smallest directions into steps that are numerically meaningless.
SVEN~\cite{Bright-Thonney:2026tlr} inverts $J$ only on its $k_{\rm SV}$ leading singular directions and discards any direction below $10^{-12}$ of the largest. Figure~\ref{fig:app:optimizer} shows the effect from one initial spectrum at $\dmax=4$: gradient descent and $k_{\rm SV}\leq8$ stall many orders of magnitude above the threshold, while $k_{\rm SV}=16$ and $32$ reach $\mathcal L\sim1$ within $150$ iterations, with $32$ no worse than $16$. The production searches use $k_{\rm SV}=8(\dmax-2)$, a fresh sample $\calT$ at every iteration, and a fixed sample of $952$ modular parameters for every reported loss.

\emphb{Cumulative Operator Count.}  Loss alone does not say whether an admitted spectrum resembles a CFT at that point. It is possible that the spectrum obtained via minimization may greatly differ from that of known CFTs. This is in fact the case to some degree. If we choose the exact same $\dgap$ and $c$ as a known CFT, the optimized spectrum will not necessarily agree with the known spectrum in the truncated region as the loss function contains several flat directions of nearly degenerate minima.
Additionally, known spectra typically have some primaries with degeneracy greater than 1; this can be approximated in the primal method by grouping operators with scaling dimensions that are sufficiently close, but disagreement remains for the same reason.

\begin{figure}[t]
    \centering
    \includegraphics[width=0.98\linewidth]{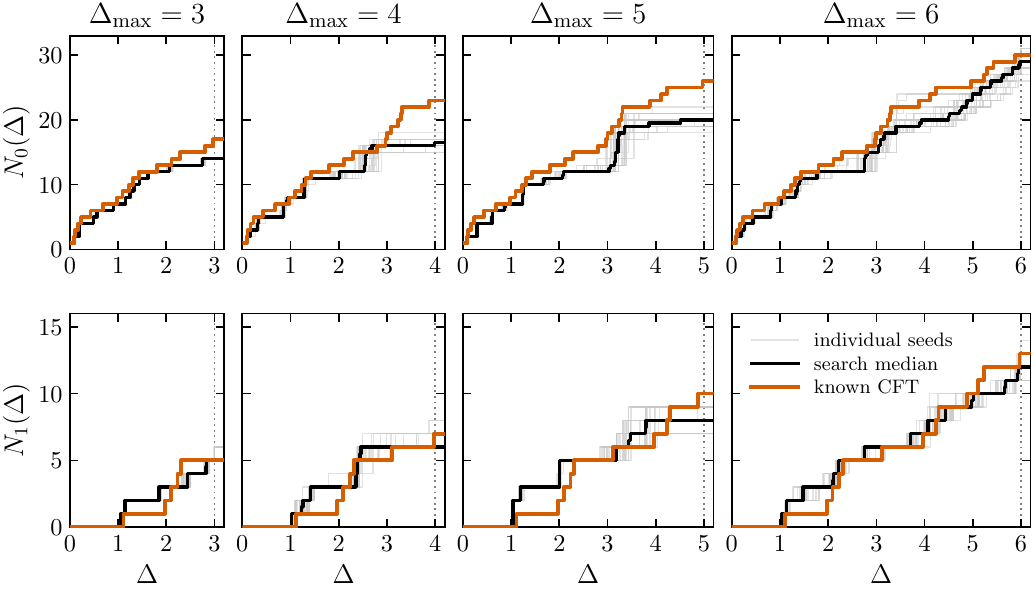}
    \caption{Cumulative number of scalar primaries $N_0(\Delta)$ (top, vacuum included) and of spin-$1$ pairs $N_1(\Delta)$ (bottom) below dimension $\Delta$ for the $\mathcal N=1$, $m=5$ minimal model at $(c,\dgap)=(81/70,3/35)$: 
    the known spectrum (orange), the median over $32$ optimization seeds (black) and the individual seeds (gray), for $\dmax=3$, $4$, $5$, and $6$ (dotted line). At $\dmax=6$ the search reproduces the known counts to within one or two operators; at lower $\dmax$ it misses scalars in the last unit below the cutoff and places a spin-$1$ pair too low.}
    \label{fig:app:cumulative_count}
\end{figure}

However, we can compare the cumulative number of operators below a certain scaling dimension; this tends to agree rather well between known CFTs and spectra obtained via optimization and also doesn't suffer an ambiguity in defining degeneracy.
We define the cumulative counts:
\begin{equation} \label{eq:supp_cumulative_counts}
N_0(\Delta)=1+\sum_{0<\Delta'\leq\Delta}d_{\Delta',0},\qquad
 N_1(\Delta)=\sum_{\Delta'\leq\Delta}d_{\Delta',1},
\end{equation}
where the $1$ is the vacuum and $N_1$ counts opposite-spin pairs.
Figure~\ref{fig:app:cumulative_count} compares spectral content obtained via optimization directly with the known spectrum of the $\mathcal N=1$ minimal model with $m=5$ at $(c,\dgap)=(81/70,3/35)$. Thirty-two searches from different initial spectra all reach $\mathcal L_{\rm eval}<25$; at $\dmax=6$ their median reproduces the known counts fairly well over the whole range, while at lower $\dmax$ the last unit below the cutoff is under-populated. Additionally, one spin-$1$ pair is placed at too low a scaling dimension. The search therefore recovers the scaling of the counting of the spectrum quite well.

\begin{figure}[t]
    \centering
    \includegraphics[width=\linewidth]{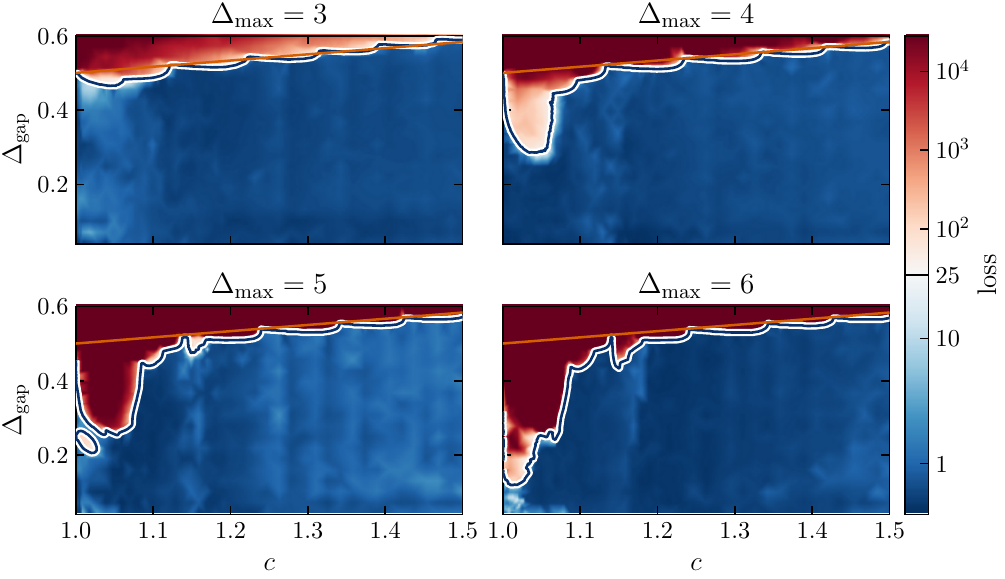}
    \caption{Evaluation loss $\mathcal L_{\rm eval}$ of the spectra found by the search at $\dmax=3$, $4$, $5$, and $6$, as a function of $c$ and $\dgap$, on a common color scale with white at $25$. The navy contour is the admission boundary $\mathcal L_{\rm eval}=25$ and the orange line is $c/6+1/3$. Away from $c=1$ the boundary is already in place at $\dmax=3$; near $c=1$ the excluded region grows with increasing $\dmax$.}
    \label{fig:app:landscapes_dmax}
\end{figure}

\begin{figure}[t]
    \centering
    \includegraphics[width=\linewidth]{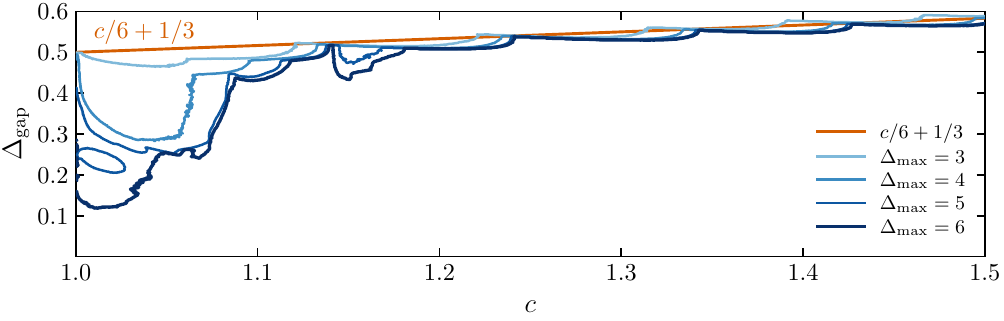}
    \caption{Boundary of the admitted region, $\mathcal L_{\rm eval}=25$, for $\dmax=3,\, 4,\,5$, and $6$, together with $c/6+1/3$. The excluded region grows monotonically with every increase of $\dmax$. }
    \label{fig:app:dmax_frontier}
\end{figure}

\emphb{The Landscape and its Dependence on $\dmax$.} Figure~\ref{fig:app:landscapes_dmax} shows the evaluation loss of the spectra found by the search over the $(c,\dgap)$ plane at $\dmax=3$, $4$, $5$, and $6$, and \Fig{app:dmax_frontier} collects the four admission boundaries. For larger $c$, the boundary is already in place at $\dmax=3$ and improves only slightly with increasing $\dmax$. Near $c=1$ the picture is different and each bound improves with increasing $\dmax$. The boundaries at different $\dmax$ are not guaranteed to be nested, since each $\dmax$ measures the residual of its spectrum against its own truncation uncertainty; what the sequence shows is where the search has converged in $\dmax$ and where it has not. Near $c=1$ it has not, and the boundary drawn in \Fig{All_Bounds} there is the $\dmax=6$ limit of the search rather than a converged bound. Going beyond $\dmax=6$ requires more than double precision, because the characters at the largest $\tau_2$ in the sample then span more decades than a double can resolve; we leave this to future work.

\begin{figure}[t]
    \centering
    \includegraphics[width=\linewidth]{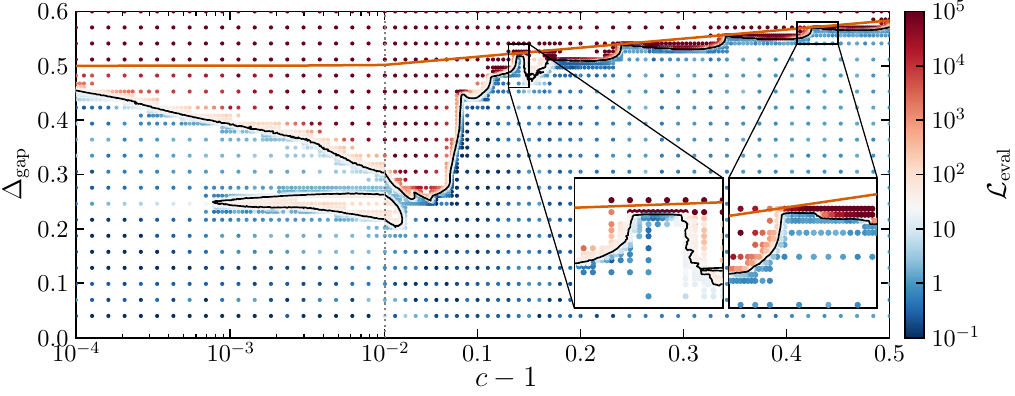}
    \caption{Sampling of the $(c,\dgap)$ plane in the $\dmax=5$ search: every dot is one evaluated spectrum, colored by its loss $\mathcal L_{\rm eval}$, with $c-1$ on a logarithmic axis below $10^{-2}$ and a linear one above. A coarse grid covers the whole plane and is refined only where the loss crosses the admission threshold $\mathcal L_{\rm eval}=25$ (black line), so the sampled cells crowd along the boundary and resolve the excluded island near $c-1\sim10^{-3}$. The insets enlarge the boxed regions $c\in[1.13,1.15]$, $\dgap\in[0.46,0.54]$ (left) and $c\in[1.41,1.45]$, $\dgap\in[0.54,0.58]$ (right). The orange line is $c/6+1/3$.}
    \label{fig:app:mesh}
\end{figure}

\emphb{Sampling and Refinement.} Figure~\ref{fig:app:mesh} shows every spectrum evaluated in the $\dmax=5$ search. A coarse grid, logarithmic in $c-1$ below $10^{-2}$ and linear above, covers the plane; cells whose neighbors straddle the threshold are subdivided and the subdivision is repeated until the boundary is resolved at the drawn scale. Each new cell is searched from several warm starts seeded by admitted spectra at neighboring cells as well as from cold starts, and is called excluded only when every start fails. The primal method chooses a specific value of $\dgap$ such that the boundary obtained is not strictly an upper bound and need not be simply connected, such as the island below the main boundary near $c-1\sim10^{-3}$. The drawn boundary is the level set $\mathcal L_{\rm eval}=25$ of the interpolated loss over the sampled cells.

\section{Agreement Between Primal and Dual Methods}

\begin{figure}[t]
    \centering
    \includegraphics[width=\linewidth]{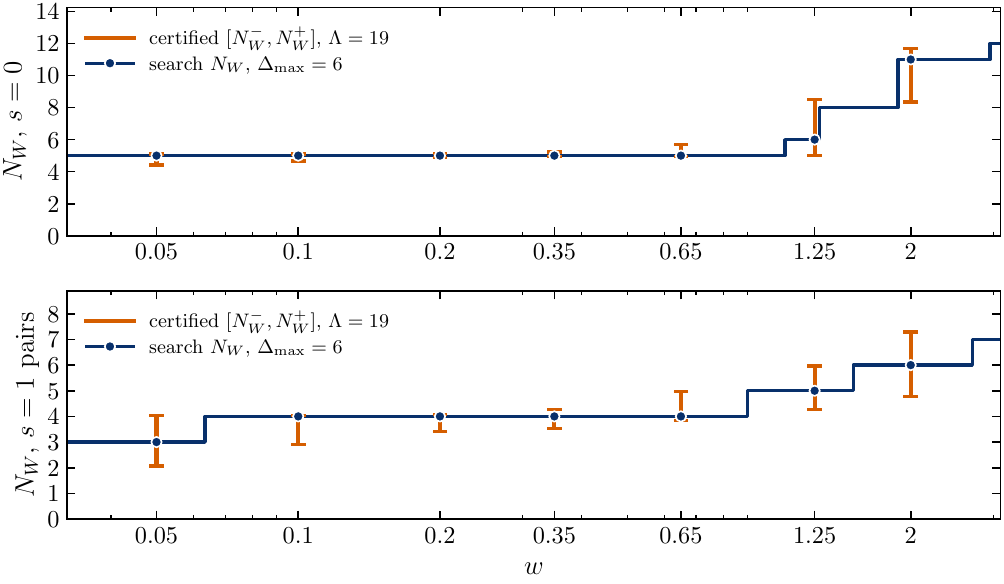}
    \caption{Number of scalars $N_W$ in the window $[\dgap,\dgap+w]$ (top) and of spin-$1$ parity pairs in $[1,1+w]$ (bottom) for the spectrum found by the search at $c=1.14$, $\dgap=0.517$ (navy, $\dmax=6$), against the certified bracket $[N_W^-,N_W^+]$ of the dual method at $\Lambda=19$ (orange), as a function of the window width $w$. Every bracket contains the search count.}
    \label{fig:app:count_containment}
\end{figure}

\begin{figure}[h!]
    \centering
    \includegraphics[width=\linewidth]{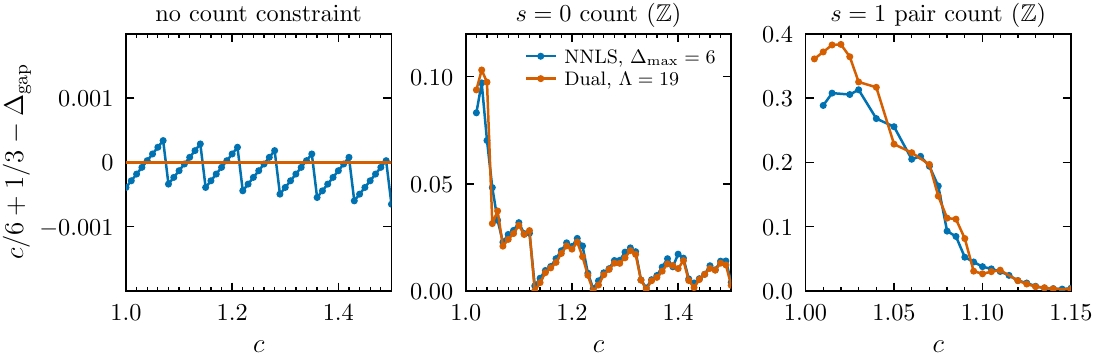}
    \caption{The continuous relaxation of the search against the certified dual bounds at $\Lambda=19$, plotted in every panel as the distance below the density bound $c/6+1/3$. Left: with no count constraint, the
    NNLS relaxation at $\dmax=6$ saturates the density bound; the residual sawtooth is the gap grid of the bisection. Middle: scalar sector, where a single integer count row reproduces the teeth of the dual scalar-count bound. Every dual point is a certified exclusion; at $c=1.02$, $1.03$ and $1.05$ the onset was not bracketed from below, so the $\Lambda=19$ bound there may be stronger than drawn. Right: spin-1 sector on the charges where both computations exist, where the NNLS relaxation tracks the dual $s=1$ bound with parity pair count restricted to $\mathbb Z$.}
    \label{fig:app:relaxation}
\end{figure}

The primal and dual methods can be compared directly wherever a spectrum found by the search sits at a central charge where count bounds have been certified. We compare the count of a candidate spectrum obtained by optimization using the primal method in a window against the dual bracket for that window. We also relax the search to impose only what the dual method imposes. We do so by developing a slightly different search method where the optimization is over a convex function. We compare the bounds on $\dgap$ obtained via this method against the dual gap bounds and largely find agreement.

\emphb{Window Counts of a Candidate Spectrum.}
Figure~\ref{fig:app:count_containment} takes the primal spectrum found at $c=1.14$, $\dgap=0.517$, the peak of the search boundary in \Fig{All_Bounds}, and counts its scalars in the windows $[\dgap,\dgap+w]$ and its spin-1 parity pairs in $[1,1+w]$. For every window, the count lies inside the dual bracket $[N_W^-,N_W^+]$ at $\Lambda=19$, computed at the same $c$ and $\dgap$. The spectrum sits where the two boundaries touch, and its integer content is exactly what the linear functionals allow.

\emphb{A Continuous Relaxation of the Search.} The residual of \Eq{primal_loss} is linear in the degeneracies. If the dimensions are fixed on a fine grid in $(\Delta,s)$ and only the weights $d\geq0$ are varied, with the vacuum and one scalar at $\dgap$ held fixed, the loss becomes:
\begin{equation}
 \mathcal L=\frac{1}{|\calT|}\,|Md-b|^2,\qquad
 M_{\tau,(\Delta,s)}=\frac{F_{\Delta,s}(\tau)}{\sigma_{\rm trunc}(\tau)},\qquad
 b(\tau)=-\frac{F_{\rm vac}(\tau)+F_{\dgap,0}(\tau)}{\sigma_{\rm trunc}(\tau)},
 \label{eq:supp_nnls}
\end{equation}
a non-negative least-squares problem (NNLS), which is convex and can be solved to its global minimum. This relaxation assumes only what the dual method assumes, a non-negative density of primaries, and can be compared with it directly. A window count can be imposed by one additional row, $\lambda\,(u_W^{T}d-n)$, with $u_W$ selecting the grid points in the window and $n$ the integer tested. Figure~\ref{fig:app:relaxation} shows the result at $\dmax=6$. With no count constraint, the relaxation of the primal method to continuous degeneracies admits every gap up to $c/6+1/3$; small deviations from $c/6+1/3$ represent finite grid size effects. With one integer count row in the scalar sector, it reproduces the teeth of the dual scalar-count bound at $\Lambda=19$, and in the spin-$1$ sector, with an integer count of parity pairs, it largely tracks the dual spin-$1$ bound for $c\gtrsim1.03$; near $c=1$ the primal and dual methods are not expected to converge for spin-1 window searches since both bounds still grow significantly with different cutoffs, $\dmax$ and $\Lambda$.

\section{Why is spin 1 interval the best?} 
\label{sec:spin1}
We get by far the strongest bounds on $\dgap$ by imposing integrality in the spin-$1$ sector, using the strategy outlined in the main text.
A heuristic explanation for why we believe spin $1$ gives the most powerful bounds is the following.
When the central charge $c$ is extremely close to $1$, we can approximate its spectrum with a $c=1$ free boson.
At generic radius $r$, the spin-$1$ primaries consist of two currents of dimension $1$ and four vertex operators with dimension $\frac12\left(r^2+r^{-2}\right)$.
Thus, when the radius $r$ becomes very large, there is a large ``desert" in energy where there are no spin-$1$ operators.
Indeed, in our numerics, we find that at the optimal $w$ the excluded bracket $[N_W^-,N_W^+]$ lies strictly between $3$ and $4$ parity pairs in $\left[1,1+w\right]$.
By contrast, the $c=1$ free boson has in its spin-$0$ sector a vacuum state $(\Delta=0)$ and a first excited state at $\Delta = \frac12\text{min}(r^2,r^{-2})$.
We see that unlike at spin $1$, this cannot be made arbitrarily large.
As a result, the scalar bounds give far  weaker improvements in $\dgap$ near $c = 1$.

\section{Catalog of Known CFTs with $1<c<\frac32$}
\label{sec:catalog}

Here, we present a catalog of known 2d unitary, compact, bosonic CFTs with central charge between $1$ and $\frac32$.  This catalog was curated with assistance from both Claude Opus 5.5 and ChatGPT-6 Astra.

We do not claim a full classification.
The entries are built from minimal models and GKO cosets~\cite{Goddard:1984vk,Goddard:1986ee}, with field identification~\cite{Schellekens:1989uf,Fuchs:1996dd}, simple-current invariants~\cite{Schellekens:1989am,Kreuzer:1993tf}, orbifolds~\cite{Dijkgraaf:1989hb,Vafa:1986wx}, and permutation invariants from braided autoequivalences~\cite{Etingof:2009yvg,Davydov:2013pic}, each of which defines a full CFT~\cite{Fuchs:2002cm}.
A list of rational CFTs with $1<c<2.1$ appears in Ref.~\cite{Antunes:2025huk}.
All known theories, as far as we are aware, in this range of $c$ are rational, although there are proposed renormalization group flows of coupled minimal models and coupled anyon chains that give candidate irrational CFTs~\cite{Dotsenko:1998gyp,Antunes:2025erb,Antunes:2025huk}.

We find $6$ infinite families of CFTs, and $47$ additional CFTs, given in Tables \ref{tab:fam} and \ref{tab:exc} respectively. All pairs of $(c, \dgap)$ are plotted as $\times$'s in \Fig{All_Bounds}. At $c=\frac{3}{2}$, the bosonized $\mathcal N=1$ super-circle line~\cite{Dixon:1988ac} fills $0<\dgap\le \frac{3}{8}$; this is the gray bar in \Fig{All_Bounds}.

\begin{table}[h]
\centering
\begin{tabular}{@{}l@{\hspace{2em}} l @{\hspace{2em}}l@{\hspace{2em}} l@{}}
\toprule
Theory & Range & $c$ & $\Dg$ \\
\midrule
$\mathrm{Ising}\times\mathcal M_m^{(A,A)}$ & $m\ge 4$ & $\tfrac32-\tfrac{6}{m(m+1)}$ & $\tfrac{3}{2m(m+1)}$ \\[3pt]
$\mathrm{Ising}\times\mathcal M_m^{D}$ & $m\ge 5$ & same & $\tfrac{4}{m(m+1)}$ for $m\ge6$; $\tfrac18$ at $m=5$ \\[3pt]
$(\mathrm{Ising}\times\mathcal M_m)/\mathbb Z_2$ & $m\ge 4$ & same & $\tfrac{4}{m(m+1)}$ \\[3pt]
$\mathcal N_m^{(A,A)}$ & odd $m\ge 5$ & $\tfrac32-\tfrac{12}{m(m+2)}$ & $\tfrac{3}{m(m+2)}$ \\[3pt]
$\mathcal N_m^{(A,A)\pm}$ & even $m\ge 6$ & same & $\tfrac{3}{m(m+2)}$ \\[3pt]
$\mathcal N_m^{(A,D)\pm},\ \mathcal N_m^{(D,A)\pm}$ & even $m\ge 8$ & same & $\tfrac{8}{m(m+2)}$ \\
\bottomrule
\end{tabular}
\caption{The infinite families of known 2d CFTs. $\mathcal M_m$ is the Virasoro minimal model~\cite{Belavin:1984vu,Friedan:1983xq} with $c_m=1-\tfrac{6}{m(m+1)}$ (Ising $=\mathcal M_3$, TIM $=\mathcal M_4$), with its A- or D-type modular invariant~\cite{Cappelli:1986hf,Kato:1987td}; $\mathbb Z_2$ is the diagonal spin flip. $\mathcal N_m^{(G,G')\pm}$ is the bosonized $\mathcal N$=1 minimal model~\cite{Friedan:1984rv}, the coset $\mathrm{SU}(2)_{m-2}\times\mathrm{SU}(2)_2/\mathrm{SU}(2)_m$~\cite{Goddard:1986ee} with ADE invariant $(G,G')$~\cite{Cappelli:1986ed}. For even $m$, the Ramond ground state is a field-identification fixed point of the coset and resolves into two fields~\cite{Schellekens:1989uf,Fuchs:1996dd}; the sign is how they pair: $+$ keeps the Ramond ground state $(c/24,c/24)$ as a scalar, $-$ pairs it as $(c/24,c/24+1)$. At $m=6$, the four $(A,D)^\pm,(D,A)^\pm$ invariants have $\Dg=\tfrac{5}{48},\tfrac18,\tfrac{5}{48},\tfrac16$ instead of $\tfrac{8}{m(m+2)}$.}
\label{tab:fam}
\end{table} 

\begin{table}[p]
\caption{All theories outside the closed-form families of Table~\ref{tab:fam}, ordered by $c$. Notation: $\mathrm{PF}_k=\mathrm{SU}(2)_k/\mathrm{U}(1)$ with rotation $r$ ($\mathbb Z_k$) and charge conjugation $C$; $\mathcal W_3(p,p+1)=\mathrm{SU}(3)_{p-3}\times\mathrm{SU}(3)_1/\mathrm{SU}(3)_{p-2}$ with $C\colon W\to-W$; $\mathrm{TIM}^2$ has spin flips $f_1,f_2$ and exchange $s$; $G_k$ is the affine algebra at level $k$; the principal $\mathrm{SU}(2)$ is the one under which the defining representation is irreducible. Theories separated by ``$=$'' have identical torus partition functions (which does not by itself make them the same CFT). 
The column deg.\ is the number of primaries at $\Dg$; in every entry the lightest primary is a scalar, $h=\bar h=\Dg/2$.}

\label{tab:exc}
\scriptsize\setlength{\tabcolsep}{8pt}
\begin{tabular}{@{}l l l l l l@{}}
\toprule
$c$ & $\Dg$ & deg. & Theory & Construction & Ref. \\
\midrule
8/7 & 1/14 & 1 & \parbox[t]{2.15in}{\raggedright $\mathrm{PF}_5/C$\strut} & \parbox[t]{2.15in}{\raggedright orbifold of $\mathrm{PF}_k$ by charge conjugation $C$\strut} & \cite{Fateev:1985mm,Dijkgraaf:1989hb} \\
8/7 & 4/35 & 2 & \parbox[t]{2.15in}{\raggedright $\mathrm{PF}_5$\strut} & \parbox[t]{2.15in}{\raggedright $\mathbb Z_k$ parafermions $\mathrm{SU}(2)_k/\mathrm{U}(1)$, diagonal\strut} & \cite{Fateev:1985mm,Gepner:1986hr} \\
6/5 & 3/40 & 1 & \parbox[t]{2.15in}{\raggedright $\mathcal W_3(5,6)/C$\strut} & \parbox[t]{2.15in}{\raggedright orbifold by $C\colon W\to-W$\strut} & \cite{Fateev:1987vh,Dijkgraaf:1989hb} \\
6/5 & 4/45 & 2 & \parbox[t]{2.15in}{\raggedright $\mathcal W_3(5,6)$\strut} & \parbox[t]{2.15in}{\raggedright $\mathcal W_3$ minimal model, diagonal\strut} & \cite{Fateev:1987vh,Fateev:1987zh} \\
14/11 & 3/44 & 1 & \parbox[t]{2.15in}{\raggedright $\mathrm{SO}(5)_2/\mathrm{SU}(2)_{20}$\strut} & \parbox[t]{2.15in}{\raggedright coset by the principal $\mathrm{SU}(2)$, diagonal\strut} & \cite{Goddard:1984vk,Schellekens:1989uf,Fuchs:1996dd} \\
14/11 & 4/55 & 2 & \parbox[t]{2.15in}{\raggedright $\mathrm{SU}(5)_1/\mathrm{SU}(2)_{20}$\strut} & \parbox[t]{2.15in}{\raggedright coset by the principal $\mathrm{SU}(2)$, diagonal\strut} & \cite{Goddard:1984vk,Schellekens:1989uf,Fuchs:1996dd} \\
14/11 & 6/55 & 2 & \parbox[t]{2.15in}{\raggedright $(\mathrm{SU}(5)_1/\mathrm{SU}(2)_{20})/\mathbb Z_5$\strut} & \parbox[t]{2.15in}{\raggedright $\mathbb Z_5$ simple-current orbifold of the row above\strut} & \cite{Kreuzer:1993tf} \\
13/10 & 19/280 & 1 & \parbox[t]{2.15in}{\raggedright $\mathrm{E}_{7,1}\times\mathrm{E}_{7,2}/\mathrm{E}_{7,3}$\strut} & \parbox[t]{2.15in}{\raggedright diagonal coset, the $\mathcal W(E_7)$ minimal model\strut} & \cite{Bais:1987dc,Bais:1987zk,Arakawa:2018iyk} \\
4/3 & 1/12 & 1 & \parbox[t]{2.15in}{\raggedright $\mathrm{PF}_7/C$\strut} & \parbox[t]{2.15in}{\raggedright orbifold of $\mathrm{PF}_k$ by charge conjugation $C$\strut} & \cite{Fateev:1985mm,Dijkgraaf:1989hb} \\
4/3 & 2/21 & 2 & \parbox[t]{2.15in}{\raggedright $\mathrm{PF}_7$\strut} & \parbox[t]{2.15in}{\raggedright $\mathbb Z_k$ parafermions $\mathrm{SU}(2)_k/\mathrm{U}(1)$, diagonal\strut} & \cite{Fateev:1985mm,Gepner:1986hr} \\
252/185 & 12/185 & 1 & \parbox[t]{2.15in}{\raggedright $\mathrm{Sp}(6)_1/\mathrm{SU}(2)_{35}$\strut} & \parbox[t]{2.15in}{\raggedright coset by the principal $\mathrm{SU}(2)$, diagonal\strut} & \cite{Goddard:1984vk,Schellekens:1989uf} \\
256/187 & 10/187 & 1 & \parbox[t]{2.15in}{\raggedright $\mathrm{E}_{8,1}\times\mathrm{E}_{8,3}/\mathrm{E}_{8,4}$\strut} & \parbox[t]{2.15in}{\raggedright diagonal coset, the $\mathcal W(E_8)$ minimal model\strut} & \cite{Bais:1987dc,Arakawa:2018iyk} \\
256/187 & 16/187 & 1 & \parbox[t]{2.15in}{\raggedright $\mathrm{E}_{8,1}\times\mathrm{E}_{8,3}/\mathrm{E}_{8,4}^{\,\pi}$\strut} & \parbox[t]{2.15in}{\raggedright same coset with the exceptional permutation of $(\mathrm{E}_8)_4$; equivalently $\mathrm{E}_{8,3}/\mathrm{Spin}(16)_3$ with the two spinors exchanged\strut} & \cite{Gannon:1995jm} \\
7/5 & 3/40 & 1 & \parbox[t]{2.15in}{\raggedright $\mathrm{Sym}^2(\mathrm{TIM})$ $=$ $\mathrm{TIM}^2/D_4$ $=$ $\mathrm{PF}_8/C$\strut} & \parbox[t]{2.15in}{\raggedright orbifold of $\mathrm{TIM}^2$ by the exchange $s$\strut} & \cite{Klemm:1990df,Borisov:1997nc} \\
7/5 & 3/40 & 2 & \parbox[t]{2.15in}{\raggedright $\mathrm{TIM}^2$\strut} & \parbox[t]{2.15in}{\raggedright two copies of the tricritical Ising model\strut} & \cite{Friedan:1984rv} \\
7/5 & 7/80 & 1 & \parbox[t]{2.15in}{\raggedright $\mathrm{PF}_8/C'$\strut} & \parbox[t]{2.15in}{\raggedright orbifold of $\mathrm{PF}_8$ by the other reflection class, $Cr$\strut} & \cite{Fateev:1985mm,Dijkgraaf:1989hb} \\
7/5 & 7/80 & 2 & \parbox[t]{2.15in}{\raggedright $\mathrm{TIM}^2/V_4$ $=$ $\mathrm{PF}_8$\strut} & \parbox[t]{2.15in}{\raggedright orbifold of $\mathrm{TIM}^2$ by $\langle f_1f_2,s\rangle$\strut} & \cite{Fateev:1985mm,Dijkgraaf:1989hb} \\
7/5 & 1/8 & 2 & \parbox[t]{2.15in}{\raggedright $\mathrm{TIM}^2/V_4^{\rm tors}$ $=$ $\mathrm{TIM}^2/\mathbb Z_4$ $=$ $\mathrm{TIM}^2/D_4^{\rm tors}$ $=$ $\mathcal N_{10}^{(A,E_6)}$\strut} & \parbox[t]{2.15in}{\raggedright orbifold of $\mathrm{TIM}^2$ by $\langle f_1f_2,s\rangle$ with discrete torsion\strut} & \cite{Vafa:1986wx,Cappelli:1986ed} \\
7/5 & 3/20 & 1 & \parbox[t]{2.15in}{\raggedright $\mathrm{PF}_8/V_4'$ $=$ $\mathrm{TIM}^{2}$ with Fibonacci swap\strut} & \parbox[t]{2.15in}{\raggedright orbifold of $\mathrm{PF}_8$ by $\langle r^4,Cr\rangle\simeq\mathbb Z_2^2$; equivalently $\mathrm{TIM}^2$ with its two Fibonacci factors exchanged\strut} & \cite{Dijkgraaf:1989hb,Etingof:2009yvg,Fuchs:2002cm} \\
7/5 & 3/20 & 2 & \parbox[t]{2.15in}{\raggedright $\mathrm{TIM}^2/\mathbb Z_2^{\rm d}$ $=$ $\mathcal N_{10}^{(D,E_6)}$ $=$ $\mathrm{PF}_8/\mathbb Z_2$\strut} & \parbox[t]{2.15in}{\raggedright orbifold of $\mathrm{TIM}^2$ by the diagonal spin flip $f_1f_2$\strut} & \cite{Dijkgraaf:1989hb,Cappelli:1986ed} \\
10/7 & 4/63 & 2 & \parbox[t]{2.15in}{\raggedright $\mathcal W_3(6,7)$\strut} & \parbox[t]{2.15in}{\raggedright $\mathcal W_3$ minimal model, diagonal\strut} & \cite{Fateev:1987vh,Fateev:1987zh} \\
10/7 & 4/63 & 1 & \parbox[t]{2.15in}{\raggedright $\mathcal W_3(6,7)/C$\strut} & \parbox[t]{2.15in}{\raggedright orbifold by $C\colon W\to-W$\strut} & \cite{Fateev:1987vh,Dijkgraaf:1989hb} \\
10/7 & 5/56 & 1 & \parbox[t]{2.15in}{\raggedright $\mathcal N_{12}^{(E_6,A)}$ $=$ $\mathcal W_3(6,7)/S_3$\strut} & \parbox[t]{2.15in}{\raggedright bosonized N=1 minimal model with an E-type invariant\strut} & \cite{Cappelli:1986ed} \\
10/7 & 5/56 & 1 & \parbox[t]{2.15in}{\raggedright $\mathcal W_3(6,7)^{C_a}/C$\strut} & \parbox[t]{2.15in}{\raggedright orbifold of the row above by $C$\strut} & \cite{Beltaos:2010ka,Dijkgraaf:1989hb} \\
10/7 & 1/7 & 3 & \parbox[t]{2.15in}{\raggedright $\mathcal N_{12}^{(E_6,D)}$ $=$ $\mathcal W_3(6,7)^{D}$\strut} & \parbox[t]{2.15in}{\raggedright bosonized N=1 minimal model with an E-type invariant\strut} & \cite{Cappelli:1986ed,Beltaos:2010ka} \\
10/7 & 1/7 & 1 & \parbox[t]{2.15in}{\raggedright $\mathcal W_3(6,7)^{C_a}$\strut} & \parbox[t]{2.15in}{\raggedright charge conjugation applied to the level-3 label only\strut} & \cite{Gannon:1992ty,Beltaos:2010ka} \\
16/11 & 5/88 & 1 & \parbox[t]{2.15in}{\raggedright $\mathrm{Ising}\times\mathcal M_{11}^{(A,E_6)}$\strut} & \parbox[t]{2.15in}{\raggedright $\mathrm{Ising}\times\mathcal M_{11}$ with an exceptional modular invariant of $\mathcal M_{11}$\strut} & \cite{Cappelli:1986hf,Kato:1987td} \\
16/11 & 8/99 & 2 & \parbox[t]{2.15in}{\raggedright $\mathrm{PF}_9$\strut} & \parbox[t]{2.15in}{\raggedright $\mathbb Z_k$ parafermions $\mathrm{SU}(2)_k/\mathrm{U}(1)$, diagonal\strut} & \cite{Fateev:1985mm,Gepner:1986hr} \\
16/11 & 8/99 & 1 & \parbox[t]{2.15in}{\raggedright $\mathrm{PF}_9/C$\strut} & \parbox[t]{2.15in}{\raggedright orbifold of $\mathrm{PF}_k$ by charge conjugation $C$\strut} & \cite{Fateev:1985mm,Dijkgraaf:1989hb} \\
16/11 & 1/11 & 1 & \parbox[t]{2.15in}{\raggedright $(\mathrm{Ising}\times\mathcal M_{11}^{(A,E_6)})/\mathbb Z_2$ $=$ $\mathrm{PF}_9/D_3$\strut} & \parbox[t]{2.15in}{\raggedright diagonal $\mathbb Z_2$ orbifold of the row above\strut} & \cite{Cappelli:1986hf,Dijkgraaf:1989hb} \\
16/11 & 2/11 & 4 & \parbox[t]{2.15in}{\raggedright $\mathrm{PF}_9/\mathbb Z_3$\strut} & \parbox[t]{2.15in}{\raggedright orbifold of $\mathrm{PF}_k$ by a rotation subgroup\strut} & \cite{Gepner:1986hr,Kreuzer:1993tf} \\
35/24 & 1/12 & 1 & \parbox[t]{2.15in}{\raggedright $\mathcal N_{16}^{(A,E_7)+}$ $=$ $\mathcal N_{16}^{(D,E_7)+}$\strut} & \parbox[t]{2.15in}{\raggedright bosonized N=1 minimal model with an E-type invariant ($\pm$ as in Table~\ref{tab:fam})\strut} & \cite{Cappelli:1986ed} \\
35/24 & 1/12 & 1 & \parbox[t]{2.15in}{\raggedright $\mathcal N_{16}^{(A,E_7)-}$ $=$ $\mathcal N_{16}^{(D,E_7)-}$\strut} & \parbox[t]{2.15in}{\raggedright bosonized N=1 minimal model with an E-type invariant ($\pm$ as in Table~\ref{tab:fam})\strut} & \cite{Cappelli:1986ed,Hsieh:2020uwb} \\
19/13 & 5/104 & 1 & \parbox[t]{2.15in}{\raggedright $\mathrm{Ising}\times\mathcal M_{12}^{(E_6,A)}$\strut} & \parbox[t]{2.15in}{\raggedright $\mathrm{Ising}\times\mathcal M_{12}$ with an exceptional modular invariant of $\mathcal M_{12}$\strut} & \cite{Cappelli:1986hf,Kato:1987td} \\
19/13 & 1/13 & 1 & \parbox[t]{2.15in}{\raggedright $(\mathrm{Ising}\times\mathcal M_{12}^{(E_6,A)})/\mathbb Z_2$\strut} & \parbox[t]{2.15in}{\raggedright diagonal $\mathbb Z_2$ orbifold of the row above\strut} & \cite{Cappelli:1986hf,Dijkgraaf:1989hb} \\
19/13 & 2/13 & 1 & \parbox[t]{2.15in}{\raggedright $\mathcal X_{12}$\strut} & \parbox[t]{2.15in}{\raggedright permutation invariant of $\mathrm{Ising}\times\mathcal M_{12}^{(E_6,A)}$, exchanging Ising with a second Ising factor hidden in the $E_6$-extended algebra\strut} & \cite{Cappelli:1986hf,Etingof:2009yvg,Fuchs:2002cm} \\
22/15 & 1/15 & 1 & \parbox[t]{2.15in}{\raggedright $\mathcal N_{18}^{(E_7,A)+}$ $=$ $\mathcal N_{18}^{(E_7,D)+}$\strut} & \parbox[t]{2.15in}{\raggedright bosonized N=1 minimal model with an E-type invariant ($\pm$ as in Table~\ref{tab:fam})\strut} & \cite{Cappelli:1986ed,Hsieh:2020uwb} \\
22/15 & 1/15 & 1 & \parbox[t]{2.15in}{\raggedright $\mathcal N_{18}^{(E_7,A)-}$ $=$ $\mathcal N_{18}^{(E_7,D)-}$\strut} & \parbox[t]{2.15in}{\raggedright bosonized N=1 minimal model with an E-type invariant ($\pm$ as in Table~\ref{tab:fam})\strut} & \cite{Cappelli:1986ed,Hsieh:2020uwb} \\
22/15 & 22/315 & 1 & \parbox[t]{2.15in}{\raggedright $\mathrm{G}_{2,1}\times\mathrm{G}_{2,2}/\mathrm{G}_{2,3}$\strut} & \parbox[t]{2.15in}{\raggedright diagonal coset\strut} & \cite{Goddard:1984vk} \\
22/15 & 1/12 & 1 & \parbox[t]{2.15in}{\raggedright $\mathrm{SU}(2)_3\times\mathrm{SU}(2)_4/\mathrm{SU}(2)_7$\strut} & \parbox[t]{2.15in}{\raggedright diagonal coset\strut} & \cite{Goddard:1984vk} \\
22/15 & 2/15 & 3 & \parbox[t]{2.15in}{\raggedright $\mathrm{SU}(2)_3\times\mathrm{SU}(3)_1/\mathrm{SU}(2)_7$\strut} & \parbox[t]{2.15in}{\raggedright same coset with the $D_4$ extension $\mathrm{SU}(2)_4\subset\mathrm{SU}(3)_1$\strut} & \cite{Goddard:1984vk,Schellekens:1986mb,Bais:1986zs} \\
151/102 & 2/51 & 1 & \parbox[t]{2.15in}{\raggedright $\mathrm{Ising}\times\mathcal M_{17}^{(A,E_7)}$\strut} & \parbox[t]{2.15in}{\raggedright $\mathrm{Ising}\times\mathcal M_{17}$ with an exceptional modular invariant of $\mathcal M_{17}$\strut} & \cite{Cappelli:1986hf,Kato:1987td} \\
169/114 & 2/57 & 1 & \parbox[t]{2.15in}{\raggedright $\mathrm{Ising}\times\mathcal M_{18}^{(E_7,A)}$\strut} & \parbox[t]{2.15in}{\raggedright $\mathrm{Ising}\times\mathcal M_{18}$ with an exceptional modular invariant of $\mathcal M_{18}$\strut} & \cite{Cappelli:1986hf,Kato:1987td} \\
52/35 & 2/35 & 1 & \parbox[t]{2.15in}{\raggedright $\mathcal N_{28}^{(A,E_8)}$ $=$ $\mathcal N_{28}^{(D,E_8)}$\strut} & \parbox[t]{2.15in}{\raggedright bosonized N=1 minimal model with an E-type invariant\strut} & \cite{Cappelli:1986ed} \\
119/80 & 1/20 & 1 & \parbox[t]{2.15in}{\raggedright $\mathcal N_{30}^{(E_8,A)}$ $=$ $\mathcal N_{30}^{(E_8,D)}$\strut} & \parbox[t]{2.15in}{\raggedright bosonized N=1 minimal model with an E-type invariant\strut} & \cite{Cappelli:1986ed} \\
433/290 & 4/145 & 1 & \parbox[t]{2.15in}{\raggedright $\mathrm{Ising}\times\mathcal M_{29}^{(A,E_8)}$\strut} & \parbox[t]{2.15in}{\raggedright $\mathrm{Ising}\times\mathcal M_{29}$ with an exceptional modular invariant of $\mathcal M_{29}$\strut} & \cite{Cappelli:1986hf,Kato:1987td} \\
463/310 & 4/155 & 1 & \parbox[t]{2.15in}{\raggedright $\mathrm{Ising}\times\mathcal M_{30}^{(E_8,A)}$\strut} & \parbox[t]{2.15in}{\raggedright $\mathrm{Ising}\times\mathcal M_{30}$ with an exceptional modular invariant of $\mathcal M_{30}$\strut} & \cite{Cappelli:1986hf,Kato:1987td} \\
\bottomrule
\end{tabular}
\end{table}

\end{document}